\documentclass[12pt]{article}
\usepackage{epsf}
\begin{document}

\newcommand{\thp}{$\Theta^+$ }

\author{Kenneth H. Hicks\\
{\small \it Department of Physics and Astronomy, Ohio University, 
Athens, OH 45701, USA}
}
\title{Experimental Search for Pentaquarks}
\maketitle

\abstract{\noindent
The experimental evidence for pentaquarks, both old and new, is 
discussed.  Constraints due to $K^+N$ scattering data from previous 
decades is first reviewed, followed by experiments with positive 
evidence and those with null results.  Finally, the problem of 
the narrow width of the \thp pentaquark is discussed, along 
with theoretical implications.
}

\section{ Introduction }

There is a well-known saying
that extraordinary claims require extraordinary proof. In the world 
of hadron spectroscopy, the discovery of a new class of hadrons 
(that cannot be described in terms of the standard classes: 3-quark 
baryons or quark-antiquark mesons) could be called an extraordinary 
claim. In fact, ten experiments claim positive evidence for a new 
hadron called the \thp (see Table 1), which has the quantum numbers 
of a pentaquark made from two up quarks, two down quarks and 
one strange antiquark (structure $uudd\bar{s}$).  However, 
no single experiment meets the criteria of extraordinary proof 
expected for such a claim.  In addition, there are a number of 
high-energy experiments that did not observe the \thp (see Table 2) 
when naive estimates suggest it should have been seen.  Hence, 
there is reasonable doubt that the \thp exists.  

There have also been searches for other pentaquarks with different 
quark-flavor composition.  Evidence for the $\Xi^{--}$ pentaquark 
$(ddss\bar{u})$ 
has been seen by one experiment \cite{na49}, but has not been 
confirmed in other experiments \cite{fischer,wa89}.  
Evidence for the $\Theta_c$, where the $\bar{c}$ quark replaces 
the $\bar{s}$ in the $\Theta^+$, has been published \cite{h1}, 
but again there is no confirmation \cite{zeus-null}. 
Little more will be said about these observations, since 
confirmation is first desired.

The purpose of this review is to provide the reader with the 
experimental facts up to the present time ($i.e.$ the end of 2004).  
However, the field is developing rapidly \cite{hyodo}, 
and this review is likely to be outdated before it is published.  

On a broader scale, 
the search for the pentaquark is an example of science at work, 
where initial evidence is peer-reviewed, published and then 
checked in all possible ways. Some claims survive and some 
fail. The fate of the \thp pentaquark is yet to be determined. 

\begin{center}
\begin{table}[ht]
\caption{Published experiments with evidence for the \thp baryon.}
\centering
\begin{tabular}{|c|l|c|c|c|c|}
\hline
Reference& Group& Reaction 	& Mass 		& Width &$\sigma$'s* \\
	&	&		& (MeV)		&(MeV)	&   \\
\hline
\cite{leps} &
LEPS	& $\gamma C \to K^+ K^- X$	& $1540\pm 10$	& $<25$	& 4.6 \\
\cite{diana} &
DIANA	& $K^+ Xe \to K^0 p X$		& $1539\pm 2$	& $<9$	& 4.4 \\
\cite{clas-d} &
CLAS	& $\gamma d \to K^+ K^- p (n)$	& $1542\pm 5$	& $<21$	&$5.2\pm 0.6^\dagger$\\
\cite{saphir} &
SAPHIR	& $\gamma d \to K^+ K^0 (n)$	& $1540\pm 6$	& $<25$	& 4.8 \\
\cite{itep} &
ITEP	& $\nu A \to K^0 p X$		& $1533\pm 5$	& $<20$	& 6.7 \\
\cite{clas-p} &
CLAS	& $\gamma p\to\pi^+ K^+K^-(n)$	& $1555\pm 10$	& $<26$	& 7.8 \\
\cite{hermes} &
HERMES	& $e^+ d \to K^0 p X$		& $1526\pm 3$	& $13\pm 9$&$\sim 5$\\
\cite{zeus} &
ZEUS	& $e^+ p \to e^+ K^0 p X$	& $1522\pm 3$	& $8\pm 4$& $\sim 5$\\
\cite{cosy} &
COSY-TOF& $p p \to K^0 p \Sigma^+$	& $1530\pm 5$	& $<18$	& 4-6  \\
\cite{svd} &
SVD	& $p A \to K^0 p X$		& $1526\pm 5$	& $<24$	& 5.6  \\
\hline
\end{tabular}
\\
$^*$ Gaussian fluctuation of the background, as $N_{peak}/\sqrt{N_{BG}}$.
This ``naive" significance may underestimate the real probability of 
a fluctuation by about 1-2 $\sigma$.\\
$^\dagger$ Further analysis of the CLAS deuterium data suggest that the 
significance of the observed peak may not be as large as indicated.
\end{table}
\end{center}

\begin{center}
\begin{table}[ht]
\caption{Published experiments with non-observation of the \thp baryon.}
\centering
\begin{tabular}{|c|l|c|c|c|}
\hline
Reference& Group& Reaction 	& Limit 	& Sensitivity? \\
	&	&		& 		&	   \\
\hline
\cite{bes} &
BES	& $e^+e^- \to J/\Psi \to \bar{\Theta}\Theta$	& $<1.1\times 10^{-5}$ B.R.	& No \cite{azimov}  \\
\cite{babar} &
BaBar	& $e^+e^- \to \Upsilon (4S) \to pK^0 X$		& $<1.0\times 10^{-4}$ B.R.	& Maybe  \\
\cite{belle} &
Belle	& $e^+e^- \to B^0\bar{B}^0 \to p\bar{p}K^0 X$	& $<2.3\times 10^{-7}$ B.R.	& No  \\
\cite{lep} &
LEP	& $e^+e^- \to Z \to pK^0 X$		& $<6.2\times 10^{-4}$ B.R.	& No?  \\
\hline
\cite{hera-b} &
HERA-B	& $p A \to K^0 p X$		& $<0.02 \times \Lambda^*$ 	& No?  \\
\cite{sphinx} &
SPHINX	& $p C \to K^0 \Theta^+ X$	& $<0.1 \times \Lambda^*$ 	& Maybe  \\
\cite{hypercp} &
HyperCP	& $p Cu \to K^0 p X$	& $<0.3\%\ K^0p$ 	& No?  \\
\cite{cdf} &
CDF	& $p \bar{p} \to K^0 p X$	& $<0.03 \times \Lambda^*$ 	& No?  \\
\cite{focus} &
FOCUS	& $\gamma BeO \to K^0 p X$	& $<0.02 \times \Sigma^*$ 	& Maybe  \\
\cite{belle-h} &
Belle	& $\pi+Si \to K^0 p X$	& $<0.02 \times \Lambda^*$ 	& Yes?  \\
\cite{phenix} &
PHENIX	& $Au+Au \to K^- \bar{n} X$	& (not given) 	& Unknown  \\
\hline
\end{tabular}
\\
\end{table}
\end{center}

\subsection{ Preliminaries }

A {\it simplified} definition of a pentaquark is a particle 
with a valence structure of four 
quarks and one antiquark.  Quantum chromodynamics (QCD) does 
not forbid multiquark particles, as long as they are colorless.
Because pentaquarks can decay (``fall apart" mode) into a 
three-quark baryon and a quark-antiquark meson, 
pentaquarks were expected \cite{jaffe,strottman} to have 
wide widths. This would be difficult to observe experimentally. 
However, some theorists \cite{lipkin,dpp} suggested that 
particular quark structures might exist with a narrow width. 
This led to renewed interest in experimental searches for 
pentaquarks.

Why is it important to know whether pentaquarks (with narrow 
widths) exist?  If they do exist, then 
we have a new multiquark system which can be used to test the 
theory of quantum chromodynamics (QCD) in the nonperturbative 
regime.  Until now, most of the effort for calculations of 
nonperturbative QCD have focused on baryons and mesons.  
Several question come to mind. How tightly bound are multiquark 
systems? How much overlap does the wavefunction of a multiquark 
particle have with the final decay state? These questions show 
that pentaquarks could provide a new testbed for QCD.  In 
particular, lattice QCD has recently produced (in the quenched 
approximation) a spectrum of baryon resonances \cite{morningstar}, 
and similar studies of pentaquarks on the lattice are 
underway (see the review by Sasaki \cite{sasaki}).

If pentaquarks with a narrow width exist, then 
we will learn more about the effective forces between quarks and 
whether lattice QCD calculations can reproduce the data.  So it 
is important to do experimental searches for pentaquarks whether 
or not you believe in any particular theory that predicts a 
given multiquark state. 

It is natural to split the effort of a review of the renewed 
interest in pentaquarks into experimental an theoretical 
aspects.  The latter is reviewed in a separate article 
that follows in the same volume \cite{goeke}, where it is 
shown that there are still many theoretical questions that 
surround the possibility of a narrow pentaquark resonance. 
But experiments must first show that (narrow) pentaquarks 
really exist.  Experiment is the focus of 
the current article.

Since there were many experiments done in previous 
decades, new searches should have advantages not previously 
available.  For example, new photoproduction facilities are now 
able to gather data for multiparticle final states at least an 
order of magnitude better than before \cite{burkert}.  This 
opens the door to new precision measurements and detection of 
weakly-produced states not seen before.  With theoretical 
guidance, it is again reasonable to look at new data and 
search for evidence of narrow pentaquark states.

\section{ Kaon Nucleon Scattering Data }

The \thp resonance (if it exists) has the same quarks as a the 
combination of a $K^+$ meson together with a neutron.  Another 
way to divide the quarks gives the combination of a $K^0$ and 
a proton. These two combinations are expected to be the primary 
decay branches of the \thp and will have equal amplitudes by 
isospin symmetry.  Reversing the process, the \thp can be 
made by putting a $K^+$ beam onto a neutron (using deuterium 
or another nuclear target) or a $K^0$ beam onto a proton. A 
$K^0$ beam is difficult to produce, but there is some data 
(see the references in \cite{arndtk0}) albeit with large 
uncertainties.  A more promising avenue is to examine $K^+d$ 
scattering.  Next we review the $K^+N$ scattering data, which 
is taken mostly from bubble-chamber experiments in the 1960's 
and 1970's.

\subsection{ Partial wave analysis }

\begin{figure}
\centerline{
\epsfxsize=15pc
\epsffile{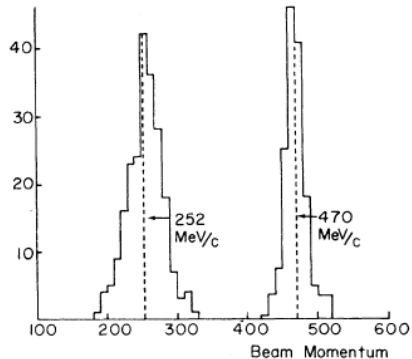}
}
\caption{ The $K^+$ beam momentum distributions for Ref. \cite{glasser}. 
Note that 470 MeV/c corresponds to a center-of-mass energy near 1550 
MeV, a bit above the average mass or the \thp in Table 1.
}
\label{fig:glasser}
\end{figure}

Let us begin with one partial wave analysis (PWA) 
of the $K^+N$ scattering done in 1992 \cite{hyslop}. In this 
paper \cite{hyslop} there is a compendium of the $K^+N$ 
scattering database.  Only some of the data has a deuteron target 
and most of these are at beam energies above about 200 MeV. 
(The average mass of the \thp shown in Table 1 is about 1535 MeV, 
which is about 100 MeV above the $K^+N$ threshold.)
Higher energy $K^+$ beams result in less $K^+$ decays on the way 
to the target, and hence typically have higher intensity and 
higher purity. Most lower-energy $K^+N$ measurements were done 
by starting with a higher-energy beam, and then decelerating 
the beam using energy loss in the target or designed energy 
degraders.  The resulting beam energy can have significant 
momentum spread, as in Fig. \ref{fig:glasser} where two 
of the four momenta measured in Ref. \cite{glasser} are shown.

\begin{figure}
\centerline{
\epsfxsize=20pc
\epsffile{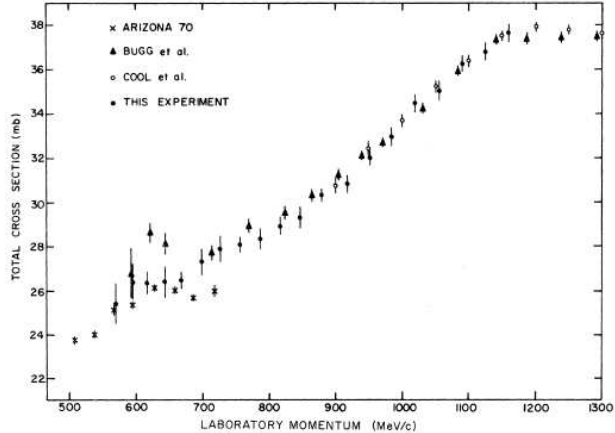}
}
\caption{ Total cross sections for $K^+d$ scattering taken 
from Ref. \cite{bowen73}.  In the region below 700 MeV/c, 
different measurements disagree significantly.
}
\label{fig:bowen}
\end{figure}

It turns out that the $K^+d$ scattering database has sparse 
coverage in the region of the \thp mass.  An additional complication 
is that there is no neutron target, and so the isospin $I=0$ 
amplitude must be extracted from deuterium data, containing a mixture of 
$I=0$ and $I=1$ amplitudes.  This can be done only after correcting 
the data for the Fermi motion of the target nucleons (and the 
momentum spread of the $K^+$ beam).  These unfolding procedures 
are straight-forward if one has sufficiently precise measurements 
at small steps in the beam energy.  An example of the Fermi 
motion correction is given in Bowen {\it et al.} \cite{bowen70} 
where structure in the energy-dependence of the 
total cross section is evident only after this procedure. 
Furthermore, the database is ``noisy" 
with significant disagreement in the regions of overlap from 
independent measurements.  This is shown in Fig. \ref{fig:bowen} 
where several measurements of the $K^+d$ total cross section 
disagree below a beam momentum of 700 MeV/c.  This is not the 
only case, as the integrated cross sections of Glasser {\it et al.} 
\cite{glasser} at 470 MeV/c (when coherent, breakup and 
charge-exchange are added together) is more than 10\% below 
the total cross sections of Bowen {\it et al.} in the same 
momentum range, a difference which is much larger than the 
systematic errors of either experiment.
Hence, one must be careful when drawing any 
conclusions from this database.

With these caveats in mind, we return to the partial wave analysis. 
In the original UCLA/Berkeley papers from the 1960's \cite{stenger} 
they conclude that the $I=0$ s-wave phase shifts (fit to their 
$K^+d$ data from 312 to 812 MeV/c) are attractive, indicating 
possible resonance structure, and for the $I=0$ p-waves there 
are two solutions.  The question of which solution to use was 
later resolved using polarization data \cite{ray} from Brookhaven, 
giving substantial attraction in the $P_{1/2}$ partial wave. 
Moving to 1977, Glasser {\it et al.} \cite{glasser} found again 
attraction in the $P_{1/2}$ partial wave, but now the s-wave 
solution was repulsive at lower beam momenta (342 and 470 MeV/c), 
turning attractive at 587 MeV/c.  Now using the full database 
as of 1993, Hyslop {\it et al.} \cite{hyslop} found that the 
s-wave is repulsive for all beam momenta, and attraction in the 
$P_{1/2}$ and also the $P_{3/2}$ partial waves (the latter was 
repulsive for Ref. \cite{stenger} and nearly zero for Ref. 
\cite{glasser}).  A more recent paper by Barnes and Swanson 
\cite{barnes} agree with Hyslop, giving a repulsive s-wave, 
but did not report the p-wave solutions. 

\begin{figure}
\centerline{
\epsfxsize=20pc
\epsffile{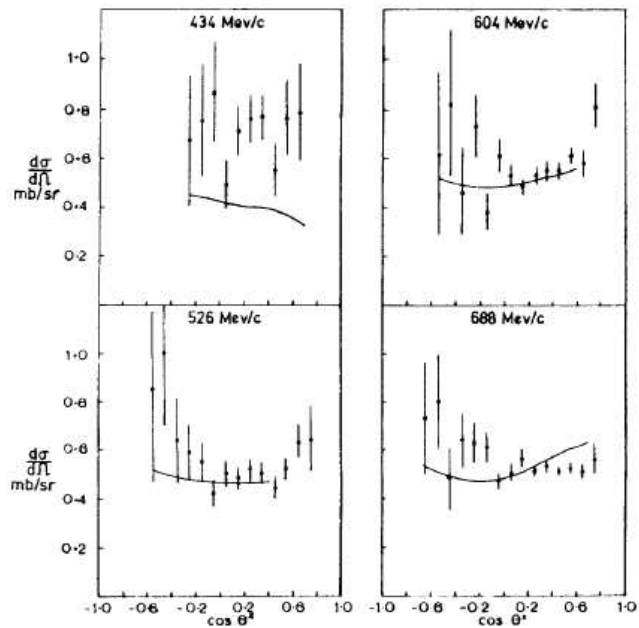}
}
\caption{ Differential cross sections for the $K^+d \to K^+n(p)$ 
reaction from Ref. \cite{damerell}.  The curves are PWA phase shift 
solutions.  Note that the 434 MeV/c data are underestimated by 
a factor of about 2.
}
\label{fig:damerell}
\end{figure}

What can we conclude from this detour into history?  
That the database is indeed noisy, and one must 
be careful to look at the individual measurements.  The 
``average" phase shifts that fit the whole database may not 
fit each data set. As an example, the $K^+d \to K^+n(p)$ 
data of Damerell {\it et al.} \cite{damerell} are shown 
in Fig. \ref{fig:damerell} for the four lowest beam momenta.
At the lowest momenta, the data are significantly underestimated 
by the overall PWA solution.  Better measurements for $K^+N$ 
scattering are needed, especially at low energy \cite{barnes}. 
It is possible that some data sets are erroneous, but we 
cannot know without better data. 

\subsection{ Connection to the \thp width }

The main point to consider here is whether the $K^+N$ database is 
consistent with the existence of a narrow \thp resonance.  Shortly 
after the first \thp papers \cite{leps}-\cite{saphir} were published, 
some comparisons with the $K^+N$ data were done.  Nussinov \cite{nussinov} 
was one of the first, and based on the general expression for the 
$K^+n$ total cross section evaluated on-resonance (with the phase 
shift at 90$^\circ$) and the momentum needed to reach the \thp mass, 
he finds a 37 mb value.  If the \thp is narrow, it could escape 
detection if there is a gap in the database at the resonant energy, 
but the deuteron's Fermi motion will spread it out so that it 
should be noticeable.  Using these estimates and a cursory 
examination of the database, Nussinov concludes that the width of 
the \thp must be less than 6 MeV.  Other estimates of the width 
followed, using similar but perhaps more careful approaches 
\cite{arndt,haidenbauer,cahn,sibirtsev,gibbs} and all of these 
agree that the \thp width must be less than a few MeV to be 
consistent with the $K^+N$ total cross section.

\begin{figure}
\centerline{
\epsfxsize=20pc
\epsffile{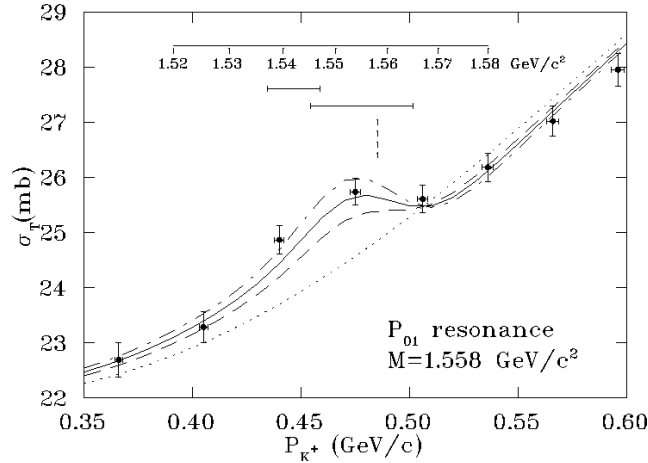}
}
\caption{ Fits to the total cross sections of Ref. \cite{bowen70} 
by W. Gibbs \cite{gibbs} using a weak-scattering approximation, 
assuming a positive-parity resonance at the mass and width shown.
}
\label{fig:gibbs}
\end{figure}

One comparison to the KN database by Gibbs \cite{gibbs} is particularly 
interesting.  This paper is based on a weak scattering approximation
and the resulting calculation is compared with the total cross section 
data of Ref. \cite{bowen70}, as shown in Fig. \ref{fig:gibbs}.  
The dotted curve is for non-resonant background, and the other 
three curves correspond to \thp widths of 0.6, 0.9 and 1.2 MeV for 
a positive parity resonance of mass about 1.56 GeV.  This resonance 
mass was obtained as the best fit to the data.  Of course, the 
uncertainties in the data allow reasonable $\chi^2$ values down to 
a mass of about 1.545 GeV.  Assuming a negative parity resonance gives 
lower mass, shown by the second horizontal bar near the top of the 
figure.  In all cases, the width of the \thp must be unusually small, 
on the order of 1 MeV. If the \thp exists with such a small width, 
then theoretical models of the quark structure of the  \thp become 
highly constrained (see section 5).

\subsection{ Possible resonant structure }

\begin{figure}
\centerline{
\epsfxsize=15pc
\epsffile{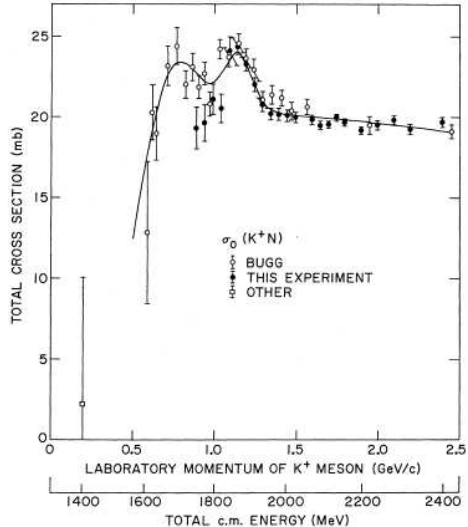}
}
\caption{ Data of Refs. \cite{cool} and \cite{bugg} for the 
isoscalar total cross section extracted from $K^+d$ scattering 
data.  The resonant-like structures at masses of about 1710 and  
1850 MeV are also identified in photoproduction \cite{tyson}.
} 
\label{fig:cool}
\end{figure}

There is stronger evidence at higher masses that there is resonant 
structure in the $K^+N$ total cross sections.  In Fig. \ref{fig:cool} 
the isoscalar cross section, $\sigma_0$ has been extracted from 
the $K^+d$ total cross section of Cool {\it et al.} \cite{cool}
and Bugg {\it et al} \cite{bugg}.  Here, there are two resonance 
structures, one at a mass of about 1710 MeV and another at about 
1860 MeV. The 1860 MeV resonance was also found in an early 
photoproduction experiment \cite{tyson} with a width of about 
150 MeV.  Higher-mass resonances were also found 
in the PWA of Hyslop {\it et al} \cite{hyslop} 
but have not received much attention.  Some experimental effort 
devoted to looking for these resonances using modern 
photoproduction facilities would make sense.

\begin{figure}
\centerline{
\epsfxsize=30pc
\epsffile{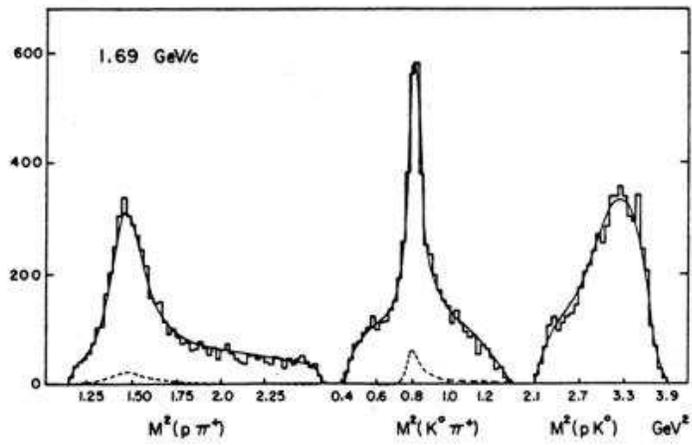}
}
\caption{ Data of Berthon {\it et al.} \cite{berthon}
for $K^+$ particles of momentum 1.69 GeV/c incident on a 
hydrogen bubble chamber.  If the pentaquark has mass 1.54 GeV 
then it would appear as a peak at $M^2 = 2.37$ GeV$^2$ in 
the $pK^0$ invariant mass spectrum on the right.
} 
\label{fig:berthon}
\end{figure}

One final comment about the KN data is from the paper by Berthon
{\it et al.} \cite{berthon} for the reaction $K^+p \to pK^0_s\pi^+$. 
This bubble chamber experiment was done at several incident kaon 
energies, with the highest momentum shown in Fig. \ref{fig:berthon}.
This figure shows several combinations of invariant mass of final 
state particles, for $M(p\pi^+)$, $M(K^0\pi^+)$ and $M(pK^0)$. 
The first shows a broad peak near the $\Delta(1232)$ mass, the 
second shows a clear peak at the mass of the $K^*(892)$ vector meson, 
and the third has a small shoulder at $M^2=2.35$ GeV$^2$ 
(or $M=1.54$ GeV).  However, further examination of the Dalitz plot 
for this reaction does not show any resonance structure, and so 
it is possible that this small shoulder in the mass distribution 
is just a statistical fluctuation.  Better data for this reaction 
is desired. An experiment at KEK \cite{imai} for the $H(K^+,\pi^+)$ 
reaction has been approved and is scheduled to run in May 2005.

\section{ Positive evidence for the \thp }

There have been many articles describing the discovery of the 
pentaquark known as the \thp at the SPring-8 facility in Japan. 
For more details about the initial discovery and two confirming 
experiments, please see Ref. \cite{nakanohicks}.

\begin{figure}
\centerline{
\hspace{3.0cm}
\epsfxsize=50pc
\epsffile{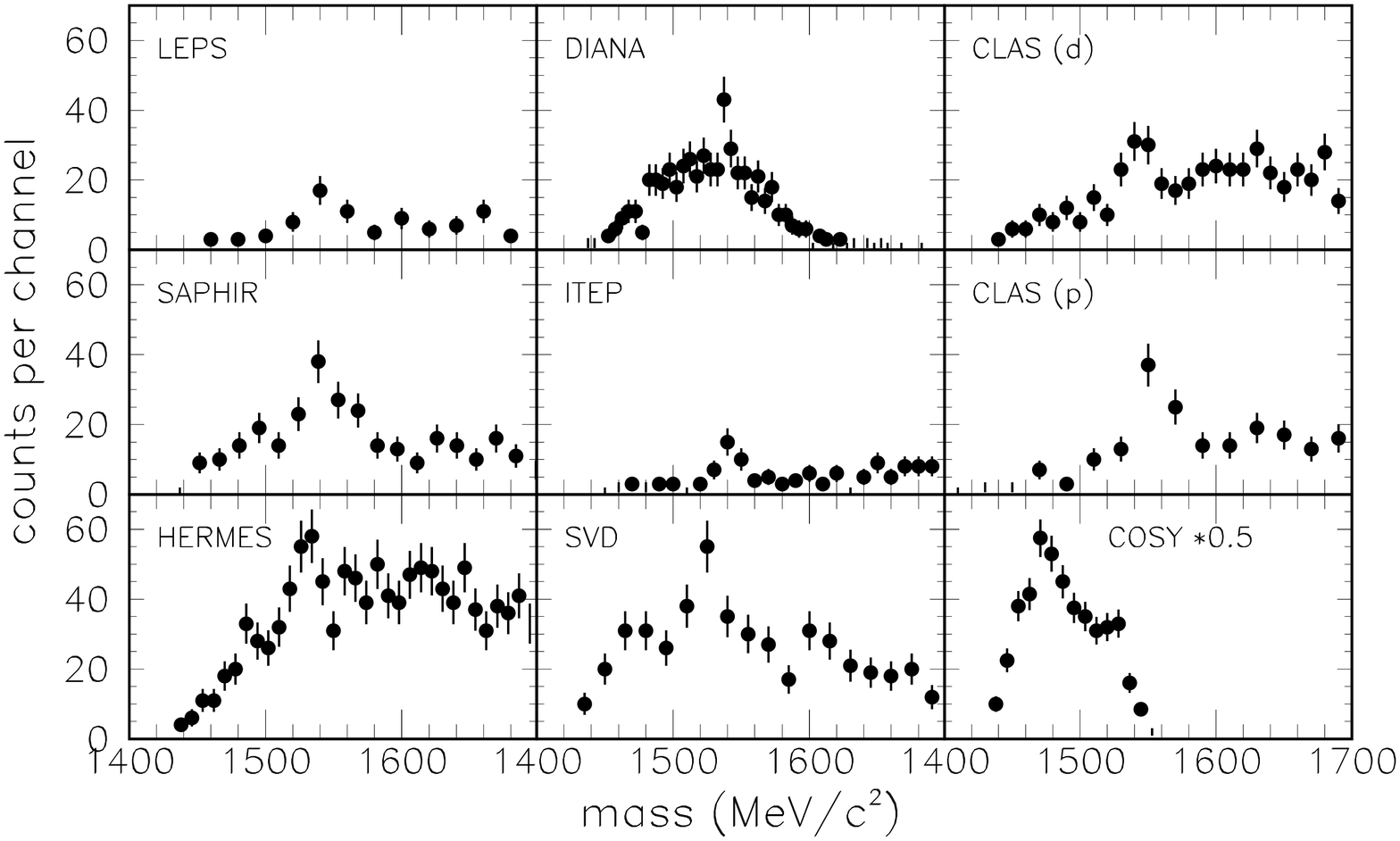}
}
\caption{ Data from experiments with positive evidence for 
the \thp (see Table 1) taken from Ref. \cite{pochod}.
} 
\label{fig:thetaall}
\end{figure}

A montage of the data from the experiments listed in 
Table 1 is shown in Fig. \ref{fig:thetaall}.  Here, the 
data are plotted with error bars and without fits to 
guide the eye.  When plotted this way, it is clear that 
better statistics are needed, since no single result 
shows a really convincing peak.  On the other hand, 
taken as a whole, there seems to be something in the 
data at 1535 MeV, independent of the probe or detector. 
Whether this might be due to coincidental statistical 
fluctuations, or some effect from the event selection 
in the analysis, is a serious question.  If we assume 
the experiments were analyzed properly then it seems 
hard to believe that so many statistical fluctuations 
could occur in the same mass, and it is tempting to 
conclude that a narrow resonance structure exists at a 
mass near 1535 MeV.  However, let us take a 
critical look at all of the individual experiments.

\subsection{ The first four experiments }

\begin{figure}
\centerline{
\epsfxsize=30pc
\epsffile{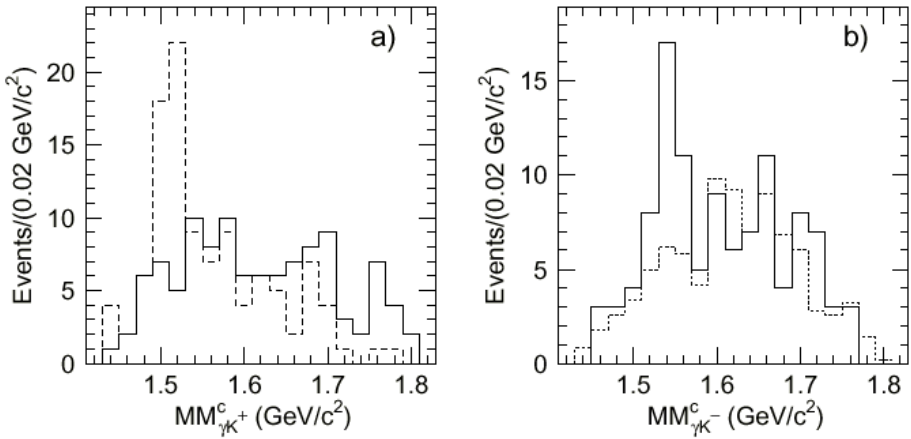}
}
\caption{ Data from LEPS \cite{leps} for the reaction 
$\gamma C \to K^+ K^- X$.  Missing mass spectra for the 
$K^+$ and $K^-$ corrected for Fermi motion are shown. 
The solid histogram shows events where no proton was 
detected. The $\Lambda(1520)$ resonance is seen on the 
left in the dotted histogram, where a coincident proton 
was detected. Possible evidence for the \thp is seen 
by the peak on the right.
} 
\label{fig:exp-leps}
\end{figure}

First results from the LEPS\cite{leps}, DIANA 
\cite{diana}, CLAS\cite{clas-d} and SAPHIR\cite{saphir} 
collaborations were ground-breaking, but each experiment 
has some weakness.  The LEPS experiment had only 19 counts 
in the peak on top of a background that was 17 counts, 
so detailed studies of the systematic uncertainties 
become very difficult.  The final plot from the 
LEPS data is shown in Fig. \ref{fig:exp-leps}, where 
the missing mass (corrected for Fermi motion) of each 
kaon in the $\gamma C \to K^+ K^- X$ reaction is 
shown, along with events where a recoil proton has 
been detected (dashed histogram).  A recoil proton 
is often detected in reactions where the proton was 
struck, such as the $\gamma p \to K^+ \ \Lambda(1520)$ 
reaction, followed by the decay $\Lambda^* \to p K^-$. 
Hence the $\Lambda(1520)$ peak is seen in the dashed 
histogram on the left. On the other hand, reactions 
on the neutron, such as $\gamma n \to K^- \ \Theta^+$ 
followed by the decay $\Theta^+ \to n K^+$, will not 
have an energetic proton. Hence the peak in the solid 
histogram on the right is interpreted as the $\Theta^+$ 
and the dashed histogram shows a possible background 
from quasi-free production of kaon pairs, which can 
happen on either protons or neutrons.  Note that the 
LEPS detector has a forward-angle-only acceptance, 
and that it is symmetric for detection of positive and 
negative particles, which is helpful when comparing 
the $\Lambda(1520)$ and $\Theta^+$ peaks.

Several questions occur when examining these data. 
First, how is the correction for Fermi motion done? 
The answer is that it is an empirical correction 
which is an approximation, good only if the momentum 
transfer to the residual nucleus is small. One 
must be careful of approximations when claiming 
evidence for a new particle.  In addition, how well 
is the background determined?  Is quasi-free production 
the dominant process, or are there other secondary 
scattering reactions that could ``reflect" into the 
phase space at the region of the peak?  If one chooses 
a different background that has a shape which is higher 
in the region of the possible $\Theta^+$ peak, then 
the evidence becomes much weaker.  Without more details 
of the proton veto efficiency and other systematic 
uncertainties in the background shape (which are not 
described in the LEPS paper \cite{leps}) then one 
must take a cautious attitude about this evidence 
for the $\Theta^+$.

\begin{figure}
\centerline{
\epsfxsize=25pc
\epsffile{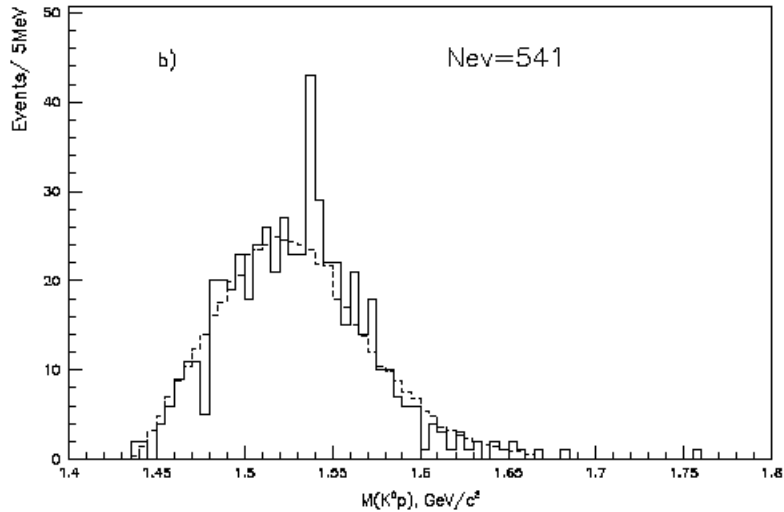}
}
\caption{ Data from DIANA \cite{diana} for the reaction  
$K^+ Xe \to p K^0_s Xe'$. The invariant mass of the $pK^0$ 
system is shown after analysis cuts (see text). The 
dotted histogram is from a mixed-event technique expected 
to represent the background.
} 
\label{fig:exp-diana}
\end{figure}

The mass spectrum from the DIANA experiment \cite{diana} 
is shown in Fig. \ref{fig:exp-diana}.  Here the reaction 
is $K^+ Xe \to K^0pXe'$ followed by the decay 
$K^0_s \to \pi^+ \pi^-$.  The final-state $p\pi^+\pi^-$ 
particles are detected by ionization tracks in the 
Xe bubble chamber photographs. The final 
data sample has cuts on the proton and $K^0$ emission 
angles (both are required to be $<100^\circ$ in the 
lab frame, and their azimuthal angles must be at 
least 90$^\circ$ apart) in order to remove rescattering 
events.  (The DIANA experiment is hampered by background 
from kaon charge-exchange reactions.) Not enough 
detail is given in their paper to show how the cuts 
they employ affect the mass 
spectrum where the \thp peak is seen, which is concentrated 
into a single bin.  In fact, if one looks at the mass 
spectrum before these angle cuts, there is only the 
slightest hint of a peak. 

One advantage of the DIANA experiment is that there is 
little chance of particle misidentification 
based on the kinematic constraints to form a $K^0$ peak 
from a $\pi^+\pi^-$ pair.  Another advantage is that the 
hadronic reaction mechanism conserves strangeness, and 
has the same quantum numbers in the initial state as for 
the $\Theta^+$.  Also, the mass spectrum is calculated 
from the invariant mass of the $pK^0$ system, and so a 
``mixed-event" technique can be used to estimate the 
background (shown by the dotted histogram in the figure) 
where protons and $K^0$'s from different events are 
combined together randomly.  This technique has been used by many 
experiments in the past as a reliable way to estimate 
the background shape.

The main disadvantage of the DIANA experiment is that 
there are no details on the sensitivity of the mass 
spectrum to the angle cuts, nor are there details on 
the modeling of the charge-exchange background (just some 
general comments that simulation studies were done \cite{diana}).  
One must be extremely cautious about peaks that appear 
only after angle cuts are made, and especially if the peak 
comes all in one bin of the histogram.  They estimate 
the statistical significance of their peak to be 4.4
$\sigma$, but a simple calculation shows that the 
single channel is only 3 standard deviations from their 
estimated background.  The combination of arbitrary 
angle cuts and small statistical significance suggests 
that this result, taken alone, is not convincing 
evidence for a new particle.

The CLAS data \cite{clas-d} was the first exclusive 
reaction, using the reaction $\gamma d \to K^+K^-p(n)$ 
on a deuterium target.  The neutron was not detected 
directly, but deduced from the missing 4-momentum, 
with very little background ($<15$\% under the 
neutron mass peak). However, detecting the proton 
requires it to have a momentum well above the 
Fermi momentum ($>300$ MeV/c) and hence the proton 
cannot be a spectator.  Since the \thp can only 
be produced (in this final state) on the neutron 
there must be a secondary reaction, such as 
rescattering of the $K^-$ from the proton, which 
complicates the reaction diagram. One might expect 
that the probability of rescattering would be low, 
but in fact it can be shown to happen with about 
30-50\% probability for $\Lambda(1520)$ production. 

\begin{figure}
\centerline{
\epsfxsize=25pc
\epsffile{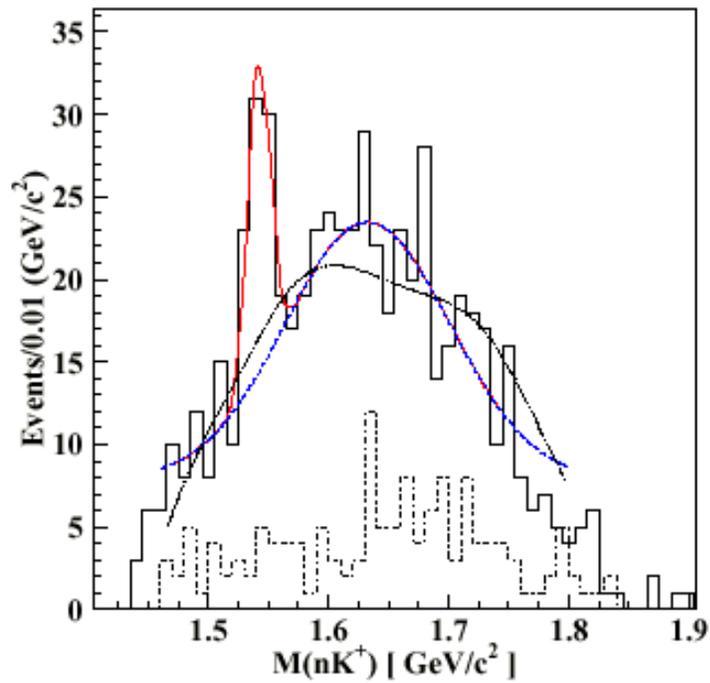}
}
\caption{ Data from CLAS \cite{clas-p} for the reaction 
$\gamma d \to K^+ K^- p (n)$. The missing mass of the 
$K^-p$ system, which is the same as the mass of the 
$nK^+$ system, is shown along with two estimates of the 
background shape (smooth curves).  The contribution of 
the $\Lambda(1520)$ events, which were cut out, are 
shown by the dashed histogram at the bottom.
} 
\label{fig:exp-clas-d}
\end{figure}

As a result, the shape of the background under the \thp 
peak is difficult to estimate and may include kinematic 
reflections\cite{dzierba}. (Note that the mechanism 
suggested by Ref. \cite{dzierba} has been challenged
because it violates C-parity conservation, see \cite{hicks}.)
However, it is true that the shape of the background 
is unknown, and the CLAS result only becomes significant 
if one accepts the background shape given in their 
paper \cite{clas-d}. An alternative background that 
stays higher, cutting through the middle of the peak 
in Fig. \ref{fig:exp-clas-d}, and then drops sharply at 
lower mass would reduce the statistical significance to 
about $3 \sigma$, rather than the 5.2 $\sigma$ for the 
background shape used in their paper.
Again, taken alone, this results could potentially be 
the result of an unlucky statistical fluctuation.

\begin{figure}
\centerline{
\epsfxsize=25pc
\epsffile{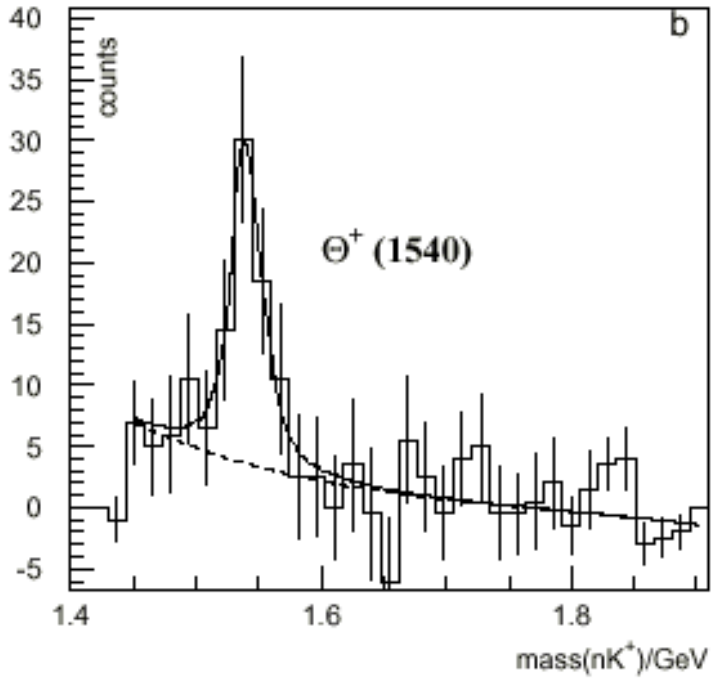}
}
\caption{ Data from SAPHIR \cite{saphir} for the reaction 
$\gamma p \to K^+ K^0_s (n)$.  The mass of the $nK^+$ 
system is shown, after sideband subtraction on either side 
of the $K^0$ peak and cuts on the $K^0$ angle (see text).
} 
\label{fig:exp-saphir}
\end{figure}

The SAPHIR collaboration was the first to publish 
results for the exclusive $\gamma p \to K_s^0 K^+ n$ 
reaction, which does not require any rescattering 
or any nuclear effects.  In principle, this is the 
best reaction to provide convincing evidence that 
the \thp exists.  Their mass spectrum is shown in 
Fig. \ref{fig:exp-saphir}, after subtraction for 
background on either side of their $K^0$ peak 
(not shown, see Ref. \cite{saphir}). In addition, 
an angle cut on the $K^0$ requiring it to be only 
at forward angles (center-of mass angle with 
$\cos\theta_{K^0} > 0.5$) has been applied.  The peak 
appears to be substantial (about 4 sigma), however  
the large cross section they estimated from their 
measurement conflicted with data for the same reaction 
measured at  CLAS\cite{hadron03}. Of course, CLAS does not 
have the same detector acceptance as SAPHIR, and if the 
\thp is produced at forward angles then SAPHIR has an 
advantage.  Still, a reanalysis of the SAPHIR 
data\cite{ostrick} suggests a smaller cross section 
(this revised analysis result has not been published).

As I write, the CLAS collaboration is preparing to release 
high-statistics results on the same reaction measured by 
SAPHIR.  The result is simply a flat mass spectrum, 
regardless of the region of the $K^0$ angle. This new 
result is in direct contradiction to the SAPHIR data. 
One can question 
whether the \thp can be produced on the proton (for 
example, the diagram given in Ref. \cite{saphir} cannot 
contribute due to C-parity violation). In fact, most
calculations \cite{ko,nam} show that production cross 
sections on the neutron are predicted to be much larger 
than those for the proton, although the calculations 
at present are very model-dependent \cite{oh}.  
In any case,  the high-statistics result of CLAS shows that 
\thp production in the $\gamma p \to K^0_s K^+ n$ 
reaction is, at the least, highly suppressed.
 
\subsection{ The $pK^0$ experiments }

Following the first reports, several experiments measured 
the invariant mass of the $K_s^0$ and a proton, from inclusive 
production.  One of these collected data from neutrino 
experiments (ITEP\cite{itep}), two others used electroproduction (
HERMES\cite{hermes} and ZEUS\cite{zeus}) and another used 
a proton beam (SVD \cite{svd}).  Of course, the $K_s^0$ 
is a mixture of both strangeness $+1$ and $-1$, so the 
invariant mass spectra will include both $\Sigma^{*+}$ and 
possible \thp peaks.  It follows that a peak at a 
mass where no $\Sigma^{*+}$ resonance is known could be 
evidence for the \thp or an unknown $\Sigma^{*+}$ resonance. 
It is also curious that these four measurements reported a  
\thp mass which is about 10 MeV below that seen by the 
first experiments (barely compatible within the 
experimental uncertainties).  The inherent weakness 
of not knowing the strangeness of a particle, 
coupled with the uncertainty in the background which must 
include the overlapping $\Sigma^{*+}$ resonances, makes 
this evidence less convincing than exclusive measurements. 
Next the details will be discussed.

\begin{figure}
\centerline{
\epsfxsize=25pc
\epsffile{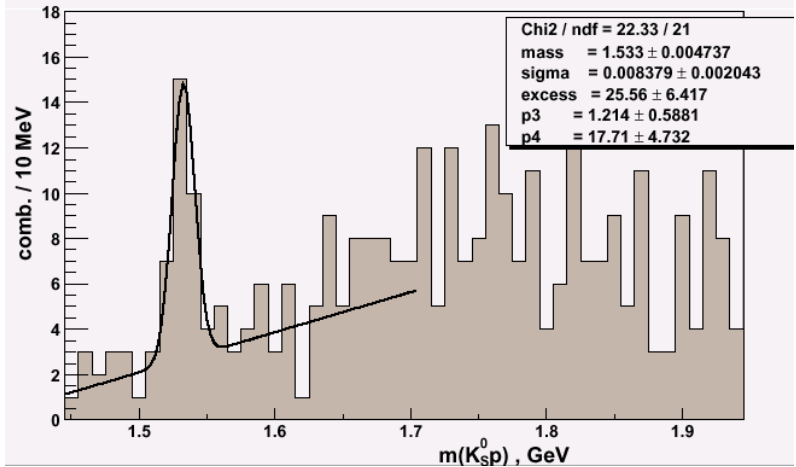}
}
\caption{ Data from ITEP \cite{itep} for the reaction 
$\nu A \to p K^0_s X$ where A represents a sum of D and Ne 
bubble chamber data.
} 
\label{fig:exp-itep}
\end{figure}

The ITEP group \cite{itep} analyzed a conglomerate of five 
neutrino experiments from bubble chambers using the 
$\nu A \to p K^0 X$ reaction, where $A$ represents either 
hydrogen (H), deuterium (D), or neon (Ne).  In these data, 
the statistics are very low for H and D, so essentially all 
of the peak shown in Fig. \ref{fig:exp-itep} comes 
from the Ne target.  (Their result is actually extracted by adding 
together the Ne and D data.)  The background can be estimated by 
taking random combinations of protons from one event and $K^0$'s 
from another event, which must be uncorrelated and hence gives a 
smooth background, which is similar to the DIANA background 
analysis. The background shown in Fig. \ref{fig:exp-itep} is just 
a linear fit to the background, whereas the random (combinatorial) 
background is a bit higher \cite{itep}.  There are about 
20 counts above a background of 12 counts, giving a 
realistic statistical significance of about 3.5 $\sigma$ 
(although they claim 6.7 $\sigma$) at a mass of 1533 MeV.

The advantages of the ITEP measurement are: (1) low background
and (2) a quantitative combinatorial background.  The disadvantages 
are: (1) the strangeness of the $K^0$ is unknown and (2) low statistics. 
In addition, there is no confirmation by modern neutrino 
experiments (although this could change in the near future), 
and a strong peak like this should clearly be visible 
at higher statistics.  Only time will tell if this peak is 
real or yet another coincidental statistical fluctuation.

\begin{figure}
\centerline{
\epsfxsize=25pc
\epsffile{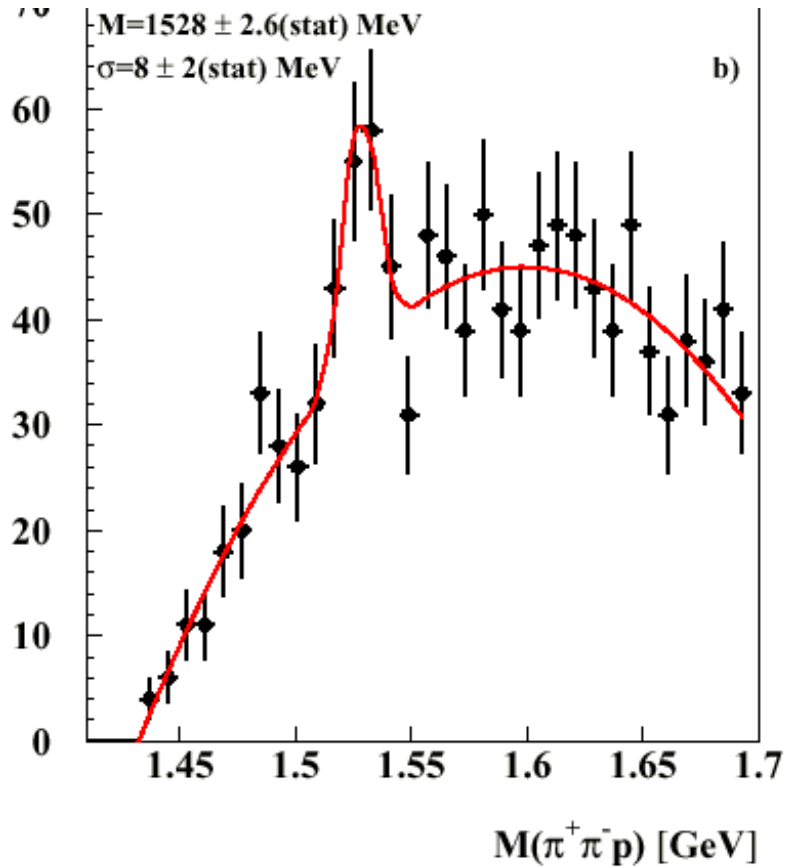}
}
\caption{ Data from HERMES \cite{hermes} for the reaction 
$e^+ d \to \pi^+ \pi^- p X$ along with a fit to the peak 
and a smooth background. 
} 
\label{fig:exp-hermes}
\end{figure}

The HERMES collaboration measured the $e^+ d \to p K^0 X$ 
reaction using the DESY accelerator positron 
beam at 27.6 GeV onto a stationary deuterium target.  Their 
event selection, based on particle identification of in a RICH 
detector \cite{hermes}, was optimized for the $pK^0$ final 
state resulting in a clear peak for the $K^0$ in the $\pi^+\pi^-$ 
invariant mass spectrum.  Combined with the proton, the $\pi^+\pi^-p$ 
invariant mass spectrum is shown in Fig. \ref{fig:exp-hermes} 
along with a fit to the peak and background.  They also tried 
other backgrounds based on simulations (using PYTHIA6) along 
with fits to known $\Sigma^{*+}$ resonances at higher mass. They 
also tried a different method for the invariant mass where the 
$K^0$ mass is used as an explicit constraint.  Depending on 
the various fits, the statistical significance of the peak 
(based on fitting errors) is between 3.4-4.3 $\sigma$. 

One of the major concerns with this result is that it is lower 
in mass at 1528 MeV with a small error (2.6 MeV). This is 
clearly inconsistent with the DIANA experiment which also 
measured the $pK^0$ invariant mass at $1539 \pm 2$ MeV.  
How can both results be the same particle, unless there are 
large systematic uncertainties in one experiment (or both)?  
A second concern is that the HERMES experiment 
does not have a strangeness tag (unlike DIANA), and so the 
peak could in principle be either the \thp or a new $\Sigma^{*+}$ 
resonance.  However, recent reanalysis of the HERMES data 
\cite{lorenzon} suggests strongly that it is unlikely to be 
$\Sigma^{*+}$ resonance. Also in Ref. \cite{lorenzon} the 
analysis was extended to remove events where an extra pion, 
not part of the $K^0$ decay, was detected and can be shown 
to be part of $\phi$ or $\Lambda$ production.  This method 
gives a better signal to background ratio, but also has 
lower statistics.  It would be interesting if more data 
could be taken, but this is unlikely to happen in the near 
future.  

\begin{figure}
\centerline{
\epsfxsize=25pc
\epsffile{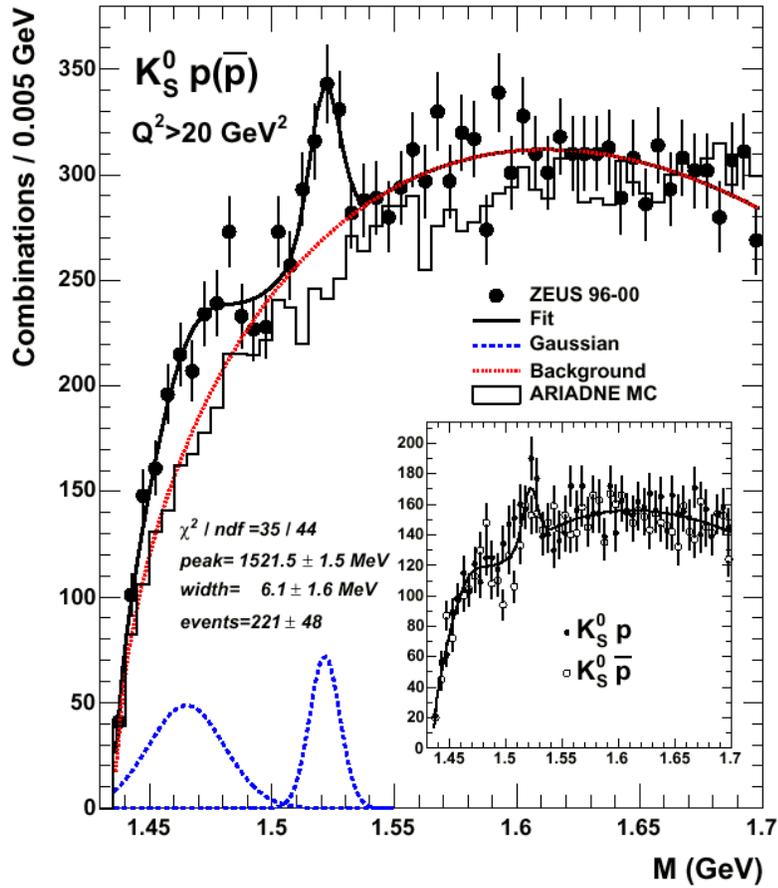}
}
\caption{ Data from ZEUS \cite{zeus} for the reaction 
$e^+ p \to K^0 p X$ with a cut on the 4-momentum transfer 
$Q^2 > 20$ GeV$^2$.  The solid line is a fit to a smooth 
background and two peaks: a known $\Sigma^{*+}$ resonance 
and a possible \thp peak at 1522 MeV. A Monte Carlo 
background is also shown by the histogram. The inset 
shows the two separate event sets added together in the 
main figure.
} 
\label{fig:exp-zeus}
\end{figure}

The ZEUS experiment \cite{zeus} did a similar measurement,
also at DESY, for the $e^+ p \to K^0 p X$ reaction except 
now at a center-of-mass energy of about 300 GeV.  At such 
high energies, hadrons are produced primarily through 
fragmentation.  The fragmentation process in the deep 
inelastic scattering (DIS) region is thought to proceed 
via ``string-breaking", where one quark is given almost 
all of the energy transfer from the scattered lepton 
followed by multiple breaking of the color-force flux-tube 
that connects the outgoing quark to the residual diquark.
As explained below, fragmentation is an unlikely method 
to produce the \thp (or any multi-quark object).
Nonetheless, the ZEUS result, shown in Fig. \ref{fig:exp-zeus} 
shows a clear peak near 1522 MeV when they take events 
with 4-momentum transfers $Q^2 > 20$ GeV$^2$. In this 
figure, both $K^0 p$ and $K^0 \bar{p}$ data have been 
added together, where the latter is for the anti-$\Theta^+$ 
since the $K^0_s$ is an equal mixture of both $K^0$ and 
anti-$K^0$ states.

The same question applies to ZEUS as it did to HERMES: 
how can the mass of their peak be consistent with the 
earlier experiments?  Of course, one could argue that 
many of those experiments had large systematic uncertainties 
on the mass, but even the ZEUS and HERMES results disagree 
in mass by 6 MeV (more than two standard deviations). 
Furthermore, why is the selection criteria $Q^2 > 20$ GeV$^2$ 
necessary?  The ZEUS data is also shown for $Q^2 > 1$ GeV$^2$ 
\cite{zeus} and no peak is visible.  These questions 
are disturbing, especially if one tries to explain away 
the null results (see below) based on suppression of 
the \thp production via fragmentation.  In other words, 
if ZEUS sees a \thp peak, then other high-energy experiments 
(where fragmentation dominates) should also see a \thp peak. 

Could the ZEUS peak possibly be due to a statistical fluctuation? 
Using simulations (with the ARIADNE package) they are able 
to estimate their background shape and then fit the 
\thp peak along with a possible lower-mass $\Sigma(1465)$ state,
giving $221 \pm 48$ events for a statistical significance of 
4.6 $\sigma$. They also did a fit without the $\Sigma^+$ state, 
getting only 3.9 $\sigma$.  In addition, they did a fit to 
the entire spectrum without {\it any} peaks, and used this 
to generate Monte Carlo spectra, and found the probability to 
make a peak, with at least 3.9 $\sigma$ somewhere in the mass 
range of 1500 to 1560 MeV, to be $6\times 10^{-5}$.  It seems 
unlikely that this peak is a statistical fluctuation.

On the other hand, it is difficult to see how the ZEUS peak 
can be the \thp unless many other experiments are wrong. 
Why is the ZEUS mass so much lower than the other experiments?  
Why is the \thp not seen in other experiments where fragmentation 
dominates?  Until these questions can be answered, 
it seems best to wait for higher statistic from ZEUS from their 
new data set (currently being analyzed).  A detailed discussion 
of the pentaquark searches at HERA are give in Ref. \cite{chekanov}.

\begin{figure}
\centerline{
\epsfxsize=25pc
\epsffile{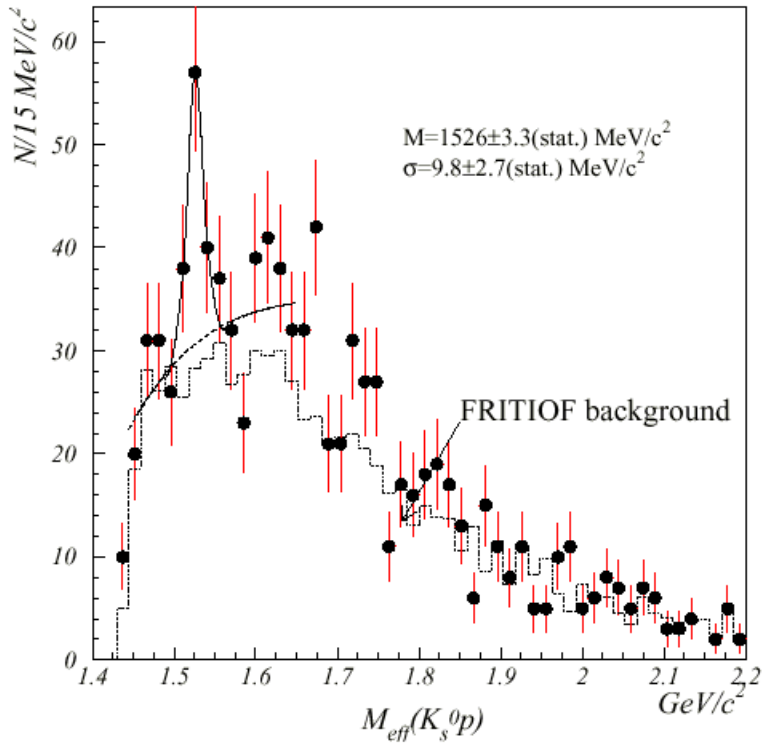}
}
\caption{ Data from SVD \cite{svd} for the reaction 
$p A \to p K^0_s X$ when the $pK^0$ system is forward of 
$90^\circ$ in the center-of-mass frame. An estimate of 
the background is shown by the FRITIOF simulation.
} 
\label{fig:exp-svd}
\end{figure}

The SVD collaboration \cite{svd} measured the $pA \to pK^0X$ 
using the 70 GeV proton beam at the IHEP accelerator in Russia. 
Their result, the invariant mass of the $pK^0$ system, is shown 
in Fig. \ref{fig:exp-svd}, using events with no more than 5 charged 
tracks and a requirement that the angle of the $pK^0$ system be 
forward of $90^\circ$ in the center-of-mass.  There is an excess 
of counts at about 1526 MeV above their background estimated 
by FRITIOF simulations. In addition, there is 
excess strength in the 1570-1750 MeV range, presumably due to 
higher-mass $\Sigma^{*+}$ resonances. Without true knowledge of the 
background, it is difficult to estimate the statistical 
significance, but it is likely about 3 $\sigma$ (or less). 

An additional cut, where the 
momentum of the $K^0$ is required to be less than the momentum of 
the proton, suppresses the $\Sigma^{*+}$ resonances (as expected based 
on kinematics of their decay) while having little effect on the 
proposed \thp peak.  With this cut, the \thp peak in their mass 
spectrum looks better \cite{svd}, but this kind of kinematic cut 
should be regarded with caution, since the kinematics of 
$\Theta^+ \to K^0p$ 
is not very different from $\Sigma^{*+} \to K^0p$ decay. 

The mass of their peak is closer to that measured by HERMES, but 
still consistent with ZEUS.  So we have the rather paradoxical 
situation that the latter four experiments, all detecting the 
$pK^0$ system with no strangeness tag, have their peaks at lower 
mass than the \thp peaks in the first four experiments, where the 
detection of the $K^+$ tags the strangeness to ensure that their spectra 
had strangeness $S=+1$.  Although it seems unlikely that there is an 
unknown $\Sigma^{*+}$ state near 1525 MeV, this possibility cannot 
be ruled out.  In any case, it is not clear that the latter four 
experiments support the case for the \thp because of this 
discrepancy in the masses of the peaks. Note that this 
is not the first time where the masses of new particles ``jumped 
around", as shown by the history plots in the Review of Particle 
Properties \cite{pdg}.  Nonetheless, the movement of the peaks 
coupled with the low statistics should make one wonder about 
the strength of the positive evidence.

\subsection{ The best positive evidence }

\begin{figure}
\centerline{
\epsfxsize=25pc
\epsffile{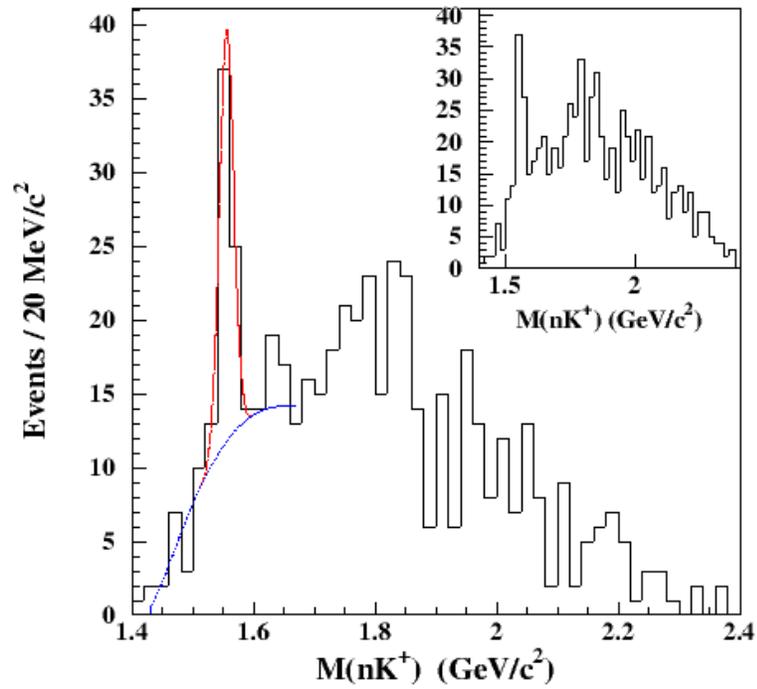}
}
\caption{ Data from CLAS \cite{clas-p} on a proton target 
for the reaction $\gamma p \to \pi^+ K^+ K^- (n)$ after 
cuts on the $K^+$ and $\pi^+$ angles. 
The inset shows the data when only the $\cos\theta_\pi > 0.8$ 
cut has been applied. The curve is a fit to the peak and 
a smooth background estimated from partial wave analysis of 
the uncut data.
} 
\label{fig:exp-clas-p}
\end{figure}

Two experiments have good evidence for the \thp.  
The first is from CLAS on a proton target \cite{clas-p}. 
This exclusive reaction, 
$\gamma p \to \pi^+ K^- K^+ n$ is very clean, and the 
background comes primarily from meson production reactions. 
The results are shown in Fig. \ref{fig:exp-clas-p} 
where cuts have been applied on the $\pi^+$ angle 
($\cos\theta_\pi > 0.8$) and $K^+$ angle 
($\cos\theta_{K^+} > 0.6$) in the center-of-mass frame.
The cuts for this analysis were not chosen arbitrary, as 
has been suggested by some critics, but are specifically 
designed to remove the dominant background (vector meson 
production) using the assumption that the \thp can be 
produced through an s-channel diagram\cite{clas-p}. 
Furthermore, these data were subjected to 
a partial wave analysis (PWA), where the amplitudes of each 
partial wave were fit over the full angular coverage of the 
CLAS detector.  Hence, the background under the \thp 
peak (after all cuts are applied) has been fixed by the 
PWA from the full (uncut) data and is represented by the 
smooth curve under the peak. 

The \thp peak here is quoted as having the highest 
statistical significance yet, in excess of 7 $\sigma$, 
although a more realistic estimate is perhaps 4 $\sigma$.  
Because this is an exclusive measurement from the proton, 
there is no ambiguity of rescattering from other nucleons, 
and the strangeness of the final state is known.  The 
authors have also checked using simulations of phase 
space, t-channel vector meson production, and data 
outside the cuts that a false peak is not artificially 
generated by the detector acceptance or from sensitivity 
to the analysis cuts. 

On the other hand, the mass of the peak is at $1.55 \pm 0.01$ GeV, 
which is about 0.01 GeV higher than the first four experiments 
and about 0.02 GeV higher than the $pK^0$ experiments.  If 
the CLAS proton result is correct, then it is incompatible 
with the HERMES, ZEUS and SVD result.  So although the CLAS 
proton results appear to be reasonably sound, the interpretation 
of the peak as the \thp is questionable since the mass is so 
different from all other experiments.

\begin{figure}
\centerline{
\epsfxsize=20pc
\epsffile{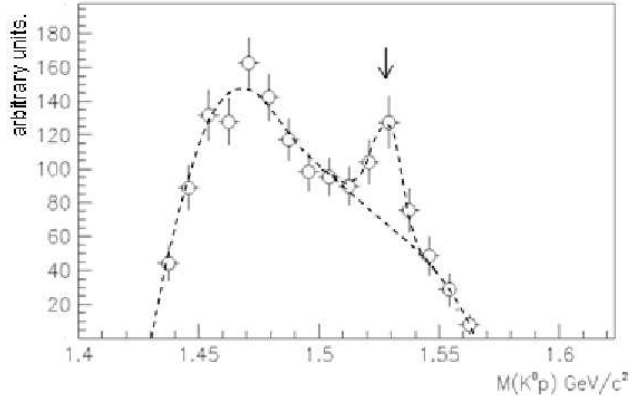}
}
\caption{ Data from COSY-TOF \cite{cosy} for the reaction 
$p p \to \Sigma^+ K^0_s p$ at near-threshold energies. 
The invariant mass of the $K^0p$ system is shown after 
acceptance corrections. 
} 
\label{fig:exp-cosy}
\end{figure}

Another experiment with good evidence for the \thp 
is the COSY-TOF result \cite{cosy} from the exclusive 
hadronic reaction $ p p \to \Sigma^+ K_s^0 p$.  
Their results are shown in Fig. \ref{fig:exp-cosy} 
which is the sum of two data sets (taken in different 
years with slightly different beam energy) after 
acceptance corrections have been applied.  Note that 
this figure is a bit different than that shown in the 
montage (Fig. \ref{fig:thetaall}) which shows the same 
data {\it before} acceptance corrections.  Here, the 
strangeness of the $pK_s^0$ invariant mass is tagged by 
the $\Sigma^+$.  The particle identification is done 
entirely by geometric reconstruction which, for this 
near-threshold reaction, is very accurate. 
Some critics have questioned whether 
this method provides good identification of the final state, 
but it can be rigorously proved that the kinematics are 
over-constrained \cite{cosy}.  The result is a very 
clean final state showing a \thp peak at a mass of 
about 1.53 GeV, which is in the middle of the \thp 
mass measurements, with a statistical significance of 
between 4 to 5 $\sigma$ (depending on the background shape).

One problem with the COSY-TOF data is that the broad hump 
near threshold (maximum at about 1.47 GeV) is unexplained. 
Could this be yet another unlucky coincidence where the 
background appears to go smoothly under the peak but in 
reality the shape is much higher in the region of the peak?
Only higher-statistics (or a quantitative calculation of 
the background) will tell us if this peak is real.  The 
COSY-TOF collaboration has taken more data in Nov.-Oct. 2004, 
which is expected to increase their statistics by a factor of 
about 5 \cite{eyrich}, with a slightly increased beam 
energy (so that the peak is not so close to the end-point 
mass.  This is really the only way to address whether the 
peak is real, by gathering much higher statistics. Until 
then, patience and caution are advised.

\section{ Non-observation experiments }

Having taken a critical look at the evidence in favor of 
the \thp we now turn to the null results.  These results 
come from either electron-positron colliders (BES\cite{bes}, 
BaBar \cite{babar}, Belle\cite{belle}, LEP\cite{lep}) 
or from high-energy reactions using a hadron 
beam (such as HERA-B \cite{hera-b}, SPHINX \cite{sphinx}, 
HyperCP \cite{hypercp} and CDF\cite{cdf}). 
Because of the difficulty in detecting neutrons in these 
detectors, these experiments look at the $pK_s^0$ 
invariant mass.  Unlike the previous experiments, the 
high-energy hadron beam experiments typically have 
higher statistics yet see no \thp peak.  

Naively, one might expect that if the \thp exists, it 
should be produced in both high-energy experiments 
through fragmentation processes as the flux tube breaks 
when the struck quark exits the nucleon. This reasoning suggests 
that the high-energy experiments should see the \thp 
(if it exists).  However, there are some curious features 
of the null results that will be discussed below.

\subsection{ The $e^+e^-$ experiments }

First, let us examine the $e^+e^-$ collider data. 
The real question here is: how can the \thp can be produced? 
The initial system starts with zero quarks, and in the 
final state, there are at least 5 quarks and 5 antiquarks.
This number of quarks is produced regularly at these 
facilities, but the reaction mechanism is thought to 
start from a quark-antiquark pair at the $e^+e^-$ annihilation 
point, and other quarks are produced in fragmentation 
or ``string breaking" as the quark-antiquark separate. 
In order to form a \thp, there must be 4 quarks and one 
antiquark localized in space with small relative velocity, 
something that would be difficult to achieve from the 
fragmentation process. A theoretical estimate by Azimov 
\cite{azimov} suggests that the BES result (see Table 2) 
is not likely to be sensitive enough to search for the \thp 
from $e^+e^-$ collisions.

\begin{figure}
\centerline{
\epsfxsize=25pc
\epsffile{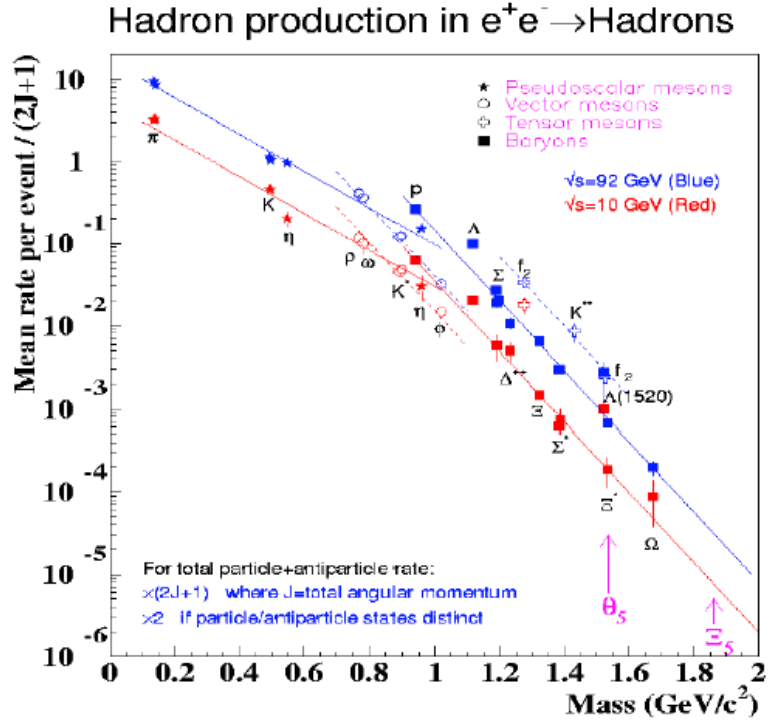}
}
\caption{ Hadron production rates from $e^+e^-$ 
collisions at BaBar \cite{babar}. The lower limits 
for \thp and $\Xi^{--}$ are shown by the arrows. 
Mesons are fit to a line at the left, and baryons 
are fit with lines having steeper slopes.  Similar 
slopes are seen for center-of-mass energies of 
10 GeV (lower lines) and 92 GeV (upper lines).
} 
\label{fig:babar-hadr}
\end{figure}

The BaBar collaboration found that the production of 
baryons follows a systematic trend 
where the probability of production decreases smoothly 
as the mass increases, see Fig. \ref{fig:babar-hadr}.
(Note: the production rate is divided by $2J+1$ where 
$J$ is the spin of the hadron.)
In this figure, it is also shown that for 
mesons, the production rate versus mass is less steep 
than for baryons.  The next question is: does the 
production rate falls off faster for pentaquark production? 
Unfortunately, there is no theory to guide us here, but 
the trend suggests that it would fall off faster, in 
which case these experiments may not have the 
sensitivity for a \thp search, although this remains
an open question. 

\begin{figure}
\centerline{
\epsfxsize=25pc
\epsffile{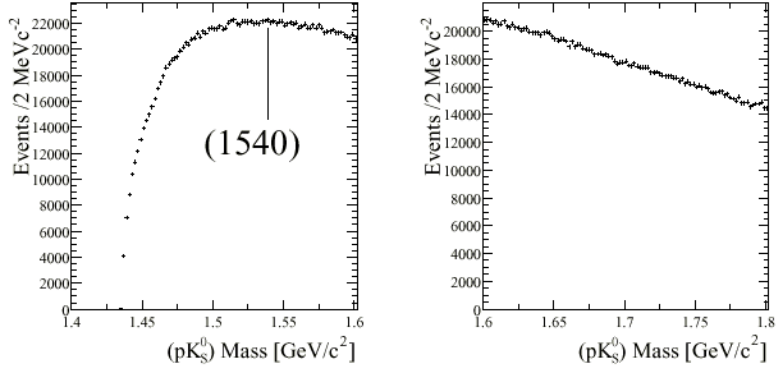}
}
\caption{ Data from BaBar \cite{babar} for inclusive 
events with a detected proton and $K^0_s$.  No 
peak is seen near the \thp mass nor are any broad 
resonances evident.
} 
\label{fig:babar}
\end{figure}

The results from the BaBar search is shown in Fig. 
\ref{fig:babar} for the invariant mass of protons and 
$K^0_s$ particles from inclusive $e^+e^-$ reactions. 
The statistics here is huge, and the resolution is 
also expected to be good ($<2$ MeV, which is the bin 
size of the points shown).  No structure is seen in 
the mass range from 1525-1560 MeV. The data can also 
be divided in to subsets for smaller ranges of the 
momentum of the proton, and again no structure is seen.
However, it is surprising that there are no structures, 
even broader peaks {\it anywhere in the mass range}! 
Since the strangeness of the $K^0_s$ is unknown, we might expect 
some $\Sigma^{*+}$ resonances (such as the $\Sigma(1670)$ which 
has a width estimated at about 60 MeV \cite{pdg}). 
In the plot to the right in Fig. \ref{fig:babar} 
the data are extremely flat, showing no signs of any 
of the known resonances.  Of course, one can argue that 
these resonances are wide and overlapping, whereas the 
\thp is expected to be quite narrow, but still it is 
surprising that there is no evidence of {\it any} 
structure, whereas the $\Lambda(1520)$ resonance (not shown) 
gives a large peak 
in the $pK^-$ invariant mass spectrum. 

The limit on pentaquark production from BaBar, given in 
Table 2, is useful information, but does not rule out 
that the \thp pentaquark could exist. Similar comments 
apply to the Belle and LEP limits from $e^+e^-$ production.
Rather, there is an opportunity here to study the rate of 
hadrons produced in $e^+e^-$ collisions, with the goal to 
elucidate the reaction mechanism and study the spectrum 
of hadron masses from these beautiful data.

\subsection{ High-energy hadron beam experiments }

Since the hadron beam experiments pose a more serious 
challenge to the existence of the \thp we should examine 
these experiments with some care.  In fairness, the same 
criticism directed at the $e^+e^-$ experiments can also be 
applied to the high-energy hadron experiments, that the 
$pK_s^0$ spectra should show evidence for known $\Sigma^{*+}$ 
resonances.

\begin{figure}
\centerline{
\epsfxsize=25pc
\epsffile{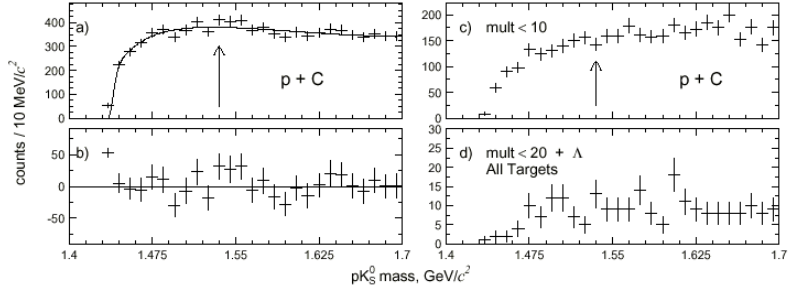}
}
\caption{ Invariant mass of a $pK^0$ pair from the 
HERA-B experiment \cite{hera-b} 
for a 920 GeV proton beam incident on a Carbon target.
The top left is for all events, and the top right is 
for a cut on the multiplicity less than 10. 
} 
\label{fig:herab}
\end{figure}

The mass spectrum from the HERA-B experiment, which 
measured the $pC \to pK^0 X$ reaction for protons at 
920 GeV onto a fixed Carbon target, is shown in 
Fig. \ref{fig:herab}. Again no peak is seen near 
1540 MeV, nor is any structure seen throughout the spectrum. 
Note that here each data point has a bin width of about 
10 MeV. Also, these data have higher particle 
multiplicities than for BaBar, which makes the combinatoric 
background (choosing which of several protons goes with the 
one or more $K^0_s$ particles) more difficult. 
The plot on the top right of Fig. \ref{fig:herab} shows the 
reduced statistics when only events with multiplicity 
$<10$ are used, which should give a ``cleaner" $pK^0$ mass
spectrum. 

\begin{figure}
\centerline{
\epsfxsize=25pc
\epsffile{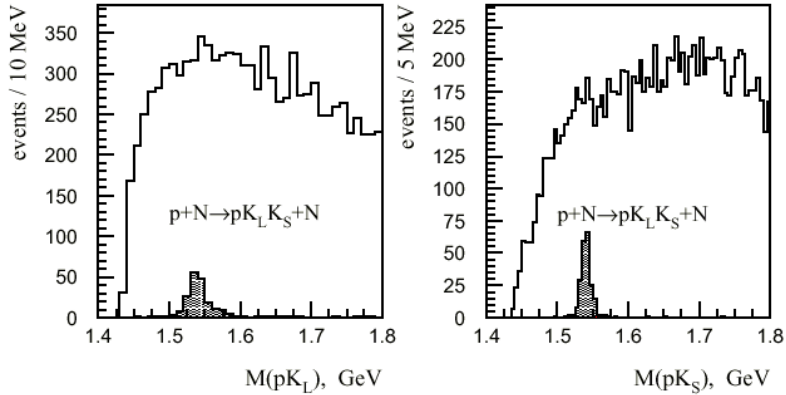}
}
\caption{ Data from SPHINX \cite{sphinx} for the reaction 
$pA \to p K_s K_L X$ at 70 GeV. The invariant mass of both 
$pK_L$ and $pK_s$ systems are shown along with a simulation 
of the \thp peak position (shaded). 
} 
\label{fig:sphinx}
\end{figure}

The SPHINX experiment measured a similar reaction, 
$pC \to p K_s K_L X$ at 70 GeV (about a factor or 10 
lower in beam energy than HERA-B), except that now 
they detected a coincident $K_L$ as a neutral cluster 
in their calorimeter. Their resulting spectra 
are shown in Fig. \ref{fig:sphinx} for the invariant 
mass of the $pK_L$ and $pK_s$ combinations, along 
with a simulation of the \thp peak resolution (shaded). 
There is no statistically significant structures, same 
as HERA-B.  The SPHINX experiment measured, at the 
same time, the $pC \to n K^+ K_s X$ where the 
neutron was detected in their calorimeter.  Again, 
there is no structure to the $nK^+$ mass spectrum. 

The HyperCP experiment at Fermilab is designed to measure 
CP violation in cascade ($\Xi$) and anti-cascade decays. 
A subset of their data was extracted for a \thp search 
from reactions on their tungsten collimator from a beam 
of mainly protons and $\pi^+$ with momenta in the range of 
100-250 GeV/c.  The experiment did not have particle 
identification.  The $K^0_s$ was reconstructed using the 
two-pion decay angles (only one pion went into their 
calorimeter).   They {\it assume} that any charged track with 
more than 50\% of the momentum is the proton.  Clearly, 
this is not an ideal setup for a pentaquark search.  (The 
authors argue that the broad momentum spread in the beam 
is an advantage, but the reader can make their own decision.)  

\begin{figure}
\centerline{
\epsfxsize=25pc
\epsffile{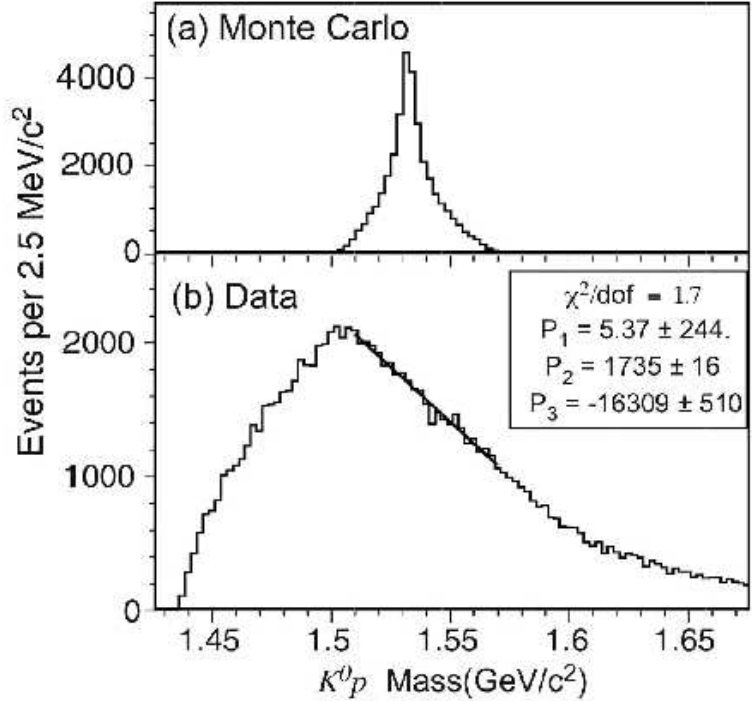}
}
\caption{ Data from HyperCP \cite{hypercp} for a mixed proton 
and pion secondary beam scattering from a tungsten collimator. 
The invariant mass of a reconstructed $K^0_s$ and proton 
candidates are shown in the lower plot, along with a simulation 
of the expected shape of a \thp peak on top. 
} 
\label{fig:hypercp}
\end{figure}

The HyperCP results are shown in Fig. \ref{fig:hypercp} 
along with a simulation showing their expected resolution 
for a \thp in their detector. 
The limit given in Table 2 assumes that the \thp is 
centered at 1.53 GeV only (no limit was given for a 
\thp at 1540 MeV, or at 1555 MeV as in the CLAS proton 
result). Note that it is possible that their spectrum could be 
contaminated with mis-identified protons that are really 
pions or kaons (since there is no particle identification).  
The above concerns suggest that one should carefully 
evaluate the usefulness of the HyperCP result. 

\begin{figure}
\centerline{
\epsfxsize=25pc
\epsffile{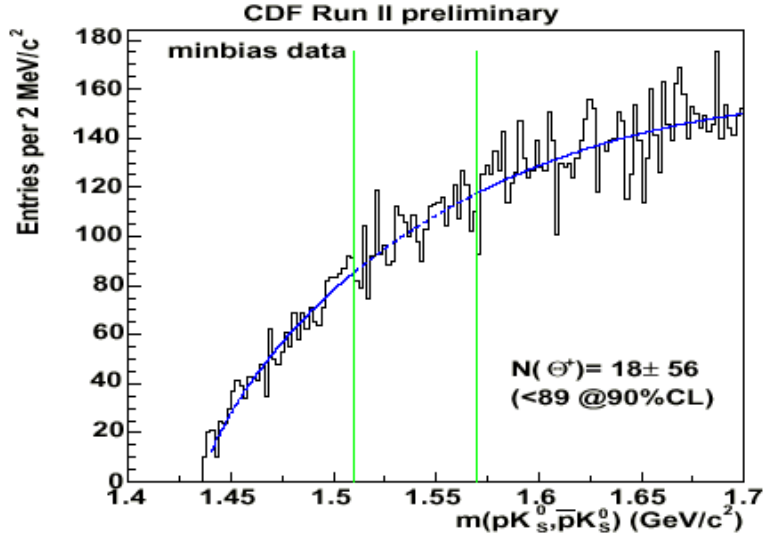}
}
\caption{ Data from CDF \cite{cdf} for the reaction 
$p \bar{p} \to p K^0 X$ for their minibias trigger. The 
shape of the background was fit to the region outside 
the horizontal lines, and an upper limit on possible 
\thp production was estimated for the range inside the 
horizontal lines.
} 
\label{fig:cdf}
\end{figure}

The CDF detector is well-known as a premier tool of high-energy 
physics.  It was good particle identification and excellent 
resolution.  They measured the $p\bar{p} \to pK^0 X$ at a 
center-of-mass energy of nearly 2 TeV.  This results in a large 
multiplicity, but the detector has the capability of resolving 
clean samples of protons and $K^0_s$'s. The invariant mass of 
the $pK^0$ system is shown in Fig. \ref{fig:cdf} with a bin 
size of 2 MeV.  A fit to the background that excludes the region 
from 1.51-1.56 GeV is shown by the solid line, and a search for 
a peak anywhere in this region was done.  Obviously, the result 
is null whereas they get several thousand counts for the 
$\Lambda(1520)$ in their $pK^-$ spectrum.  Note that the minibias 
data used here were heavily prescaled, resulting in lower 
statistics than other physics triggers.  Their paper 
\cite{cdf} does not give the multiplicity of particles in each 
event, and it would be interesting to know more details about 
the combinatoric backgrounds.

\begin{figure}
\centerline{
\epsfxsize=25pc
\epsffile{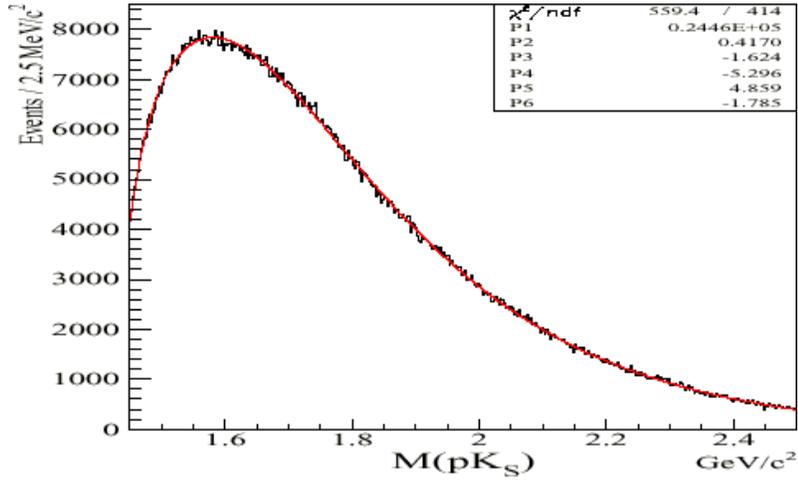}
}
\caption{ Data from the FOCUS experiment for the reaction 
$\gamma BeO \to p K^0_s X$ from a 300 GeV brehmsstrahlung 
beam.  No structures are seen here, but other channels 
such as $\Sigma^* \to \Lambda \pi$ have robust peaks.
} 
\label{fig:focus}
\end{figure}

The FOCUS experiment \cite{focus} used a photon beam from 
brehmsstrahlung of 300 GeV electrons and positrons.  The 
photons hit nuclei in a BeO target, and charged particles 
are tracked using a silicon vertex detector followed by 
momentum analysis in dipole magnets.  Particle identification 
is clean, done by three Cerenkov detectors and two EM 
calorimeters.  Their results are shown in Fig. \ref{fig:focus} 
for the invariant mass of the $pK^0_s$ system.  Few details 
are given about this analysis, such as estimates for the 
shape of this spectrum from MC simulations or why they don't 
see any $\Sigma^*$ resonances in this spectrum, but a longer 
paper with details is expected in the near future.

\begin{figure}
\centerline{
\epsfxsize=25pc
\epsffile{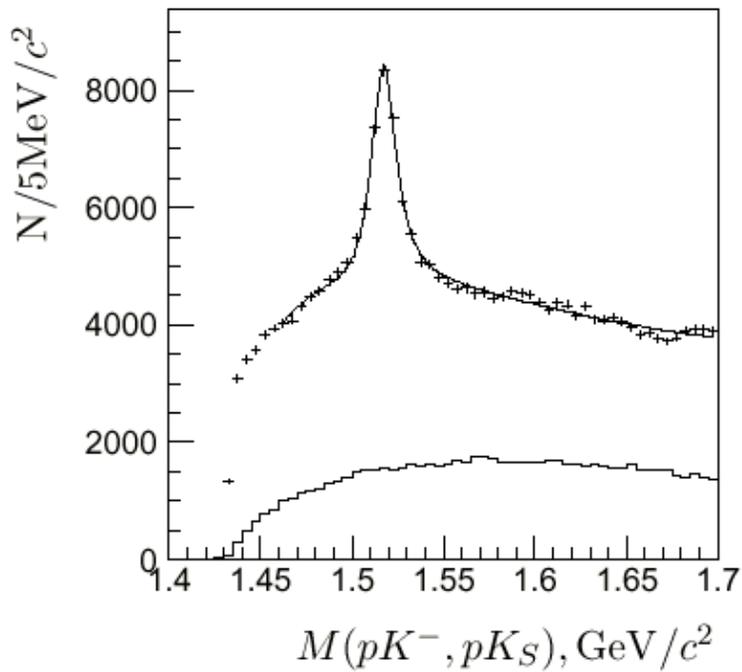}
}
\caption{ Data from Belle \cite{belle-h} for a beam of 
secondary particles (mostly protons and pions) from the 
interaction vertex scattered from their silicon vertex 
detector.  The invariant mass of the $pK^-$ system is 
shown with the $\Lambda(1520)$ peak above the spectrum 
of the $pK^0_S$ system where no structure is seen.
} 
\label{fig:belle}
\end{figure}

The Belle experiment \cite{belle-h} took a different approach.  They 
used secondary scattering of mesons (from $e^+e^-$ collisions) in 
their silicon vertex detector to produce known $Y^*$ resonances. 
If the \thp exists, it could be produced with a $K^+$ beam of the 
right energy.  Their results are shown in Fig. \ref{fig:belle}. 
Unfortunately, the hadrons incident on the silicon 
target have unknown identity and unknown energy.  Only a small 
fraction of these data could result in production of the \thp and 
detected by its decay into the $pK^0$ channel.  With the high 
resolution of Belle, even a small signal (with a narrow width) 
might be visible, but none was seen. Again, {\it we need better 
calculations of the expected number of counts (based 
on Belle's spectrum of hadrons incident on silicon)  before we 
can interpret their null result.}

\subsection{ Discussion of null results }

Based on the high-energy data, one might conclude that 
no \thp exists and also that 
no $\Sigma^*$ resonances exist.
But what is the production mechanism of the $\Sigma^*$ 
resonances?  Similarly, what is the production mechanism 
of the possible \thp resonance?
Theoretical calculations are needed in order to 
understand the true significance of the null results.

The production mechanism of the \thp (if it exists) or 
even the $\Lambda^*$ and $\Sigma^*$ resonances from 
fragmentation processes is not well known. However, 
a first step in this direction was taken by Titov {\it et al.} 
\cite{titov}, using quark constituent counting rules to 
estimate the ratio of \thp to $\Lambda(1520)$ production 
in fragmentation reactions.  Fragmentation functions are well 
established \cite{close} and have been used for years to describe 
the distribution of hadrons from high-energy collisions, based 
on the number of constituent partons in the projectile and target. 
Using this model, Titov shows that production of the $\Theta^+$ 
is suppressed relative to the $\Lambda$(1520) resonance by about 
3 orders of magnitude for experiments such as HERA-B. 
Of course, the simple model used for this estimate may not be 
a good approximation for all kinematics, but it is consistent 
with the null experimental results at high energies.

Finally, the facts should be clearly stated when drawing 
conclusions from both positive and null evidence.  The kinematics 
in the null experiments are different from those in 
the experiments reporting positive evidence.  
In other words, the null results do not prove that the 
positive results are wrong.  There may be some interesting 
physics to be learned, assuming all experiments are correct, 
as to why exclusive measurements at medium energy show a 
possible \thp peak whereas its production may be suppressed 
in high-energy inclusive measurements.  In any case, the 
onus is on the medium-energy experiments to make a better 
case for the possible existence of the $\Theta^+$.

\section{ The Problem of the Width }

Perhaps the most disturbing fact of the \thp evidence is 
that its width appears to be very narrow. Direct evidence 
\cite{diana,hermes,zeus} limits the width to be less 
than about 10 MeV.  Indirect evidence, based on analysis 
of KN scattering data \cite{nussinov,cahn,arndt,sibirtsev,gibbs}, 
estimates the width at a few MeV or less.  Such a narrow 
width for a resonance 100 MeV above its strong decay threshold 
would be unprecedented. 

Coupled with the narrow width problem is the 
question of parity.  The spin of the lowest-lying \thp 
is expected to be $J=1/2$ with either negative (S-wave) or 
positive (P-wave) parity. A narrow width from an S-wave 
resonance makes no sense \cite{jw} whereas a P-wave 
would allow a centrifugal barrier so that a narrow width 
at least possible \cite{jw,kl}.  It seems logical that 
if the \thp width is narrow, its parity must be positive. 
This idea was beautifully presented by Hosaka \cite{hosaka}.

What do lattice QCD calculations say about the parity?  
Several lattice results are known \cite{sasaki} 
and except for one result \cite{chiu}, only the negative 
parity projection gives a signal consistent with the \thp 
state.  So we have an apparent contradiction between the parity 
deduced from quark models (above) and the parity deduced from 
(most) lattice calculations.  One obvious resolution to this 
dilemma is to conclude that the \thp does not exist. 
In fact, more recent lattice studies with more CPU power 
have concluded that there is no mass eigenstate that is 
separable from the low-lying scattering states 
\cite{uklat,csikor}.  However, 
we must realize that the lattice calculations for exotic 
baryon resonances should be regarded as exploratory \cite{sasaki}. 
Extrapolating to the chiral limit from the heavy quarks 
used in lattice calculations must be done properly \cite{chiu,uklat} 
and furthermore, all lattice calculations are done in the 
quenched approximation.  Hence we should be wary of 
parity statements based on current lattice results.

Even if the \thp exists with $J^P=\frac{1}{2}^+$, a width as 
narrow as 1 MeV is theoretically difficult to understand 
\cite{prasz}.  However, several new theoretical ideas 
show that such a narrow width is consistent with theory.  
Ellis, Karliner and Praszalowicz \cite{ellis} have shown 
that, when higher-order terms from mixing of the low-lying 
baryon multiplets due to SU(3) symmetry breaking are 
included in the chiral soliton model, the \thp width could 
accidentally be very small.  Separately, using a two-state 
model, Karliner and Lipkin showed \cite{karliner} that the 
mass eigenstates of two pentaquarks ($e.g.$, mixtures 
of the Jaffe-Wilczek model and the diquark-triquark model) 
can mix, resulting in one coupling strongly to KN decay 
(with a wide width) and one decoupling (with a narrow width). 
In a different approach, using the QCD string model, 
Suganuma {\it et al.} showed \cite{suganuma} that the 
pentaquark does not just ``fall apart" as predicted by the 
quark model, but must overcome a sizable potential barrier 
to decay into a KN final state.  This results in a very 
narrow width for the \thp in their model.  In all, it is 
interesting that a narrow width of 1 MeV can be accommodated 
within the quark model, the chiral soliton model and the QCD 
string model.

Of course, the \thp does not necessarily have spin $J=1/2$. 
It is possible that it could have $J=3/2$, and this was 
investigated on the lattice by the Adelaide lattice group 
\cite{adelaide}.  Interestingly, they find attraction for 
$J^P = \frac{3}{2}^+$ but not in the negative parity 
state.  Also, they do not see any attraction for $J=1/2$ 
in either parity.  This is an interesting development, 
and could be tested experimentally, 
should the \thp turn out to be real (with a narrow width). 

Clearly, experimental information is needed before one can 
test the various ideas about the \thp width.  Proposals at 
KEK \cite{imai} and Jefferson Lab \cite{bogdan} for high 
resolution spectrometer experiments have been approved 
and will likely run in 2005.
Other facilities already mentioned (CLAS, ZEUS and COSY-TOF) 
will gather more statistics, which should enable a better 
determination of the \thp width (if it exists).  In addition to 
width measurements, we need to know the \thp parity.  This 
will likely be done at COSY-TOF  using polarized 
target and polarized beam, which has a clear theoretical 
interpretation as shown by Hanhart \cite{hanhart}. If the 
\thp exists, then we have the experimental tools to learn 
about its width and parity.

\section{ Summary }

After a long dormant period, there is again experimental 
action in the search for pentaquarks.  Currently, there 
is a lot of interest, both experimental and theoretical, 
in the subject. Are the peaks near 1535 MeV real or 
not?  If real, then we will learn a lot about QCD.
In particular, are correlations between quarks 
important in nonperturbative solutions of QCD?

The $K^+N$ data from the 1960's and 1970's does not show 
resonance structure near threshold, and this casts some 
doubt on the existence of the \thp pentaquark, unless 
it is has a narrow width of a few MeV or less. If it 
is narrow, then this makes the \thp (if it exists) 
difficult to understand in terms of current theoretical 
models.  This theoretical bias, coupled with the null 
results from high energy experiments, makes it difficult 
to believe the \thp is real.  Hence, the ball is in the 
experimental court.  If the \thp can be shown to exist, 
then the nay-sayers will be forgotten, and the theorists 
will find a way to explain its narrow width.

Hadronic physics, and non-perturbative QCD in general, 
may hold surprises for us.  We should seek ways to 
explore the edges of this terrain, and searching for 
multiquark states is one way to do this.  Regardless 
of the outcome, the search is worth the effort.

\subsection*{Acknowledgements}
I am indebted to many colleagues, both experimental and 
theoretical, with whom I have had discussions on this topic.
Special thanks go to Takashi Nakano, leader of the LEPS 
collaboration, and Stepan Stepanyan, co-spokesman of the 
CLAS deuterium experiment, whom are my close experimental 
partners.

\newcommand{\etal}{ {\it et al.}, }

\end{document}